\documentclass[twocolumn,english,aps,prb,groupedaddress,showpacs]{revtex4}
\usepackage[latin9]{inputenc}
\usepackage[active]{srcltx}
\usepackage{color}
\usepackage{amsmath}
\usepackage{amssymb}
\usepackage{graphicx}

\makeatletter

\@ifundefined{textcolor}{}
{%
 \definecolor{BLACK}{gray}{0}
 \definecolor{WHITE}{gray}{1}
 \definecolor{RED}{rgb}{1,0,0}
 \definecolor{GREEN}{rgb}{0,1,0}
 \definecolor{BLUE}{rgb}{0,0,1}
 \definecolor{CYAN}{cmyk}{1,0,0,0}
 \definecolor{MAGENTA}{cmyk}{0,1,0,0}
 \definecolor{YELLOW}{cmyk}{0,0,1,0}
}

\usepackage{graphicx}% Include figure files
\usepackage{dcolumn}% Align table columns on decimal point
\usepackage{bm}% bold math
\usepackage{amsfonts}
\usepackage{amsthm}
\usepackage{verbatim}
\usepackage{pstricks}

\usepackage{color}
\usepackage{epsfig}

\newcommand{\be}{\begin{equation}}
\newcommand{\ee}{\end{equation}}
\newcommand{\bea}{\begin{eqnarray}}
\newcommand{\eea}{\end{eqnarray}}

\makeatother

\usepackage{babel}
\begin{document}

\title{Long range alternating spin current order in a quantum wire with
modulated spin-orbit interactions}

\author{G.\ L.\ Rossini }

\affiliation{IFLP-CONICET and Departamento de F\'{i}sica, Universidad Nacional
de La Plata, CC 67 1900 La Plata, Argentina }

\author{D.\ C.\ Cabra}

\affiliation{IFLySiB-CONICET and Departamento de F\'{i}sica, Universidad Nacional
de La Plata, CC 67 1900 La Plata, Argentina}

\author{G.\ I.\ Japaridze}

\affiliation{Faculty of Natural Sciences and Medicine, Ilia State University,
Tbilisi, Georgia}

\affiliation{Andronikashvili Institute of Physics, Tbilisi, Georgia}
\begin{abstract}
A key concept in the emerging field of spintronics is the electric
field control of spin precession via the effective magnetic field
generated by the Rashba spin orbit interaction (RSOI). Here, by extensive
Density Matrix Renormalization Group computations, we demonstrate
the presence of alternating spin current order in the gapped phases
of a quantum wire with spatially modulated RSOI and repulsive electron-electron
interactions. Our results are analytically supported by bosonization
and by a mapping to a locally rotated spin basis.
\end{abstract}


\date{\today}
\maketitle

\section{Introduction\label{sec:Introduction}}

The possibility to manipulate magnetization at nanoscale using the
coupling between the electron spin and its motion (orbital angular
momentum) has led to the emergence of a new research field named \char`\"{}spin-orbitronics\char`\"{}
{[}\onlinecite{Wolf_et_al_Science_01,Fabian_Zutic_Review_04,Fabian_Zutic_Review_09,Awschalom_etal_13}{]}.
The main advantage of this approach is based on the exploitation of
the spin-orbit (SO) interaction {[}\onlinecite{Winkler_Book_03}{]}
to get efficient ways for manipulating the magnetization in integrated
spintronic systems and create a low power storage and/or logic devices
{[}\onlinecite{Awschalom_etal_02,Morton_etal_11}{]}. The seminal
proposal of Datta and Das for a spin field-effect transistor highlights
the use of the SO interaction {[}\onlinecite{Datta_Das_90}{]}. A
basic ingredient of the Datta-Das transistor is a ballistic quantum
wire with sufficiently strong Rashba spin-orbit interaction (RSOI)
{[}\onlinecite{Rashba_60}{]}, the latter is required for creating
a sizeable spin precession. Depending on spin orientations in the
source and in the drain one can modulate the current flowing through
the device and thus implement in principle ON/OFF states. The strength
of the spin-orbit coupling can be tuned by applying a gate voltage
to the system {[}\onlinecite{DD_Transistor_1,DD_Transistor_2}{]}.
In two dimensional structures, coupling between the charge and spin
degrees of freedom via the SO interaction provides a mechanism for
efficient conversions between charge and spin currents. The spin Hall
effect {[}\onlinecite{Dyakonov_Perel_71,HE_Sinova_etal_RMP_15}{]}
by which a charge current can be converted into a transverse spin
current and the inverse spin Hall effect {[}\onlinecite{Inv_HE_1,Inv_HE_2}{]}
for the inverse conversion are the primary examples.

In last years the SO effects in quasi-one-dimensional strongly correlated
electron systems have became the subject of intensive studies due
to their fascinating properties and wide possibilities to engineer
new materials with unconventional electronic and magnetic properties.
This includes \emph{helical conductors} which appear in the presence
of strong spin-orbit interaction in quantum wires {[}\onlinecite{Streda_03}{]},
nanotubes {[}\onlinecite{Klinovaja_etal_11}{]} or on the edges of
topological insulators {[}\onlinecite{Hasan_Kane_RMP_2010}{]}. Helical
conductors have became of topical interest because their robustness
with respect to the disorder {[}\onlinecite{SJJ_10}{]} and because
they offer the possibility for spin-filtered transport {[}\onlinecite{Streda_03,Cabra_etal_17}{]},
Cooper pair splitting {[}\onlinecite{Sato_etal_10}{]} and, if in
contact with a superconductor, the realization of Majorana bound states
at their ends {[}\onlinecite{Alicea_12,Fu_Kane_08,Lutchyn_etal_10,Oreg_etal_10,Alicea_etal_1l,Yazdani_etal_13,MJJ_Paper_16}{]}.

Another fascinating property of the SO interaction is that it can
be exploited to engineer magnetic materials in which new types of
topological objects, such as chiral domain walls or magnetic skyrmions
can be stabilized (see for recent review {[}\onlinecite{Balents_Savary_17,Legrand_etal_18}{]}).
Such spin configurations are driven by an additional term in the exchange
interaction, namely Dzyaloshinskii-Moriya interaction (DMI) {[}\onlinecite{Dzyaloshinskii_57}{]},
which arises from the presence of SO coupling and inversion symmetry
breaking {[}\onlinecite{Moriya_60}{]}. In quasi-one-dimensional magnetic
materials the DMI is responsible for formation of a chiral order {[}\onlinecite{Dzyaloshinskii_64,Oshikawa_Affleck,Aristov_Maleev_00,Tsvelick_01,Starykh_08,Garate_Affleck_10,Starykh_17}{]}.
It is also the key structural element ensuring coupling between magnetic
and electric degrees of freedom in the spin-driven \emph{chiral multiferroic}
materials {[}\onlinecite{Katsura_05,Ch-MFS}{]}. These systems became
very actual in last years {[}\onlinecite{SD-Ch-MFM}{]}, in particular
in the context of materials useful for electric field controlled quantum
information processing.

Recently it has been demonstrated that the SO interaction can be tailored
with a substantial efficiency factor by external electric fields as
in a metallic phase of a quantum wire {[}\onlinecite{EF_Enhanc_SOI}{]},
as well as in the case of insulating quantum magnet {[}\onlinecite{EF_Enhanc_DMI}{]}.
This unveils the possibility to control SO interaction and magnetic
anisotropy via the electric field and opens a wide area for exploring
the effects caused by the spatially modulated SO interaction on the
properties of low-dimensional electron systems both in conducting
and insulating phases.

Theoretical studies of the one-dimensional correlated electron systems
with spatially modulated SO interactions of different genesis, counts
almost two decades {[}\onlinecite{Cabra_etal_17},\onlinecite{Mireles_Kirczenow_01,Wang_04,GongYang_07,Zhang_etal_05,Wang_etal_06,Sanchez_Serra,JJF_09,Xiao_Chen_10,MGJJ_11,JJM_14,MJJ_Paper_16,AJR_19}{]}.
A Peierls-type mechanism for a spin-based current switch was identified
in {[}\onlinecite{JJF_09}{]}, where it was shown that a spatially
smooth modulated Rashba SOI coupling opens both charge and spin gaps
in the system at commensurate band fillings. Such an interaction could
be generated by a periodic gate configuration, as sketched in {[}\onlinecite{Cabra_etal_17}{]},
or in contact with an anti-ferroelectrically ordered material {[}\onlinecite{Streltsov-20015}{]}.
In subsequent studies the effect of induced charge density wave correlations
in the quantum wire due to the periodic potential was examined, and
the optimal regime where insulating current blockade occurs was determined
{[}\onlinecite{MGJJ_11}{]}. Later it was shown that the half-metal
phase, where electrons with only a selected spin polarization exhibit
ballistic conductance, can be reached by tuning of a uniform external
magnetic field acting on a quantum wire with modulated spin-orbit
interaction {[}\onlinecite{Cabra_etal_17}{]}. More recently it was
shown that, in the case of a half-filled band and in the limit of
strong Coulomb repulsion where the charge excitations are gapped and
the spin degrees of freedom are described by an effective spin $S=1/2$
Heisenberg chain, the very presence of spatially modulated DMI substantially
enriches the ground state phase diagram of the spin system leading
to the formation of a new gapped phases with composite order characterized
by the coexisting of bond-located alternating dimerization and chirality
patterns and, for a particular parameter range, also of the staggered
on-site magnetization {[}\onlinecite{AJR_19}{]}.

In the present article we put forward studies of the insulating phases
of one-dimensional electron systems with modulated Rashba spin-orbit
interaction including, in one scheme, analysis of the band-filling
commensurability conditions necessary for the formation of \emph{band
insulating} phases {[}\onlinecite{JJF_09}{]} together with consideration
of the effects caused by the strong electron-electron interaction,
responsible for the formation of a \emph{Mott correlated insulator
phase} effectively described by the above mentioned spin chain Hamiltonian
{[}\onlinecite{AJR_19}{]}. We present a detailed study of the excitation
spectrum, as well as alternating charge and spin order in the ground
state of a one-dimensional system of electrons with spatially modulated
RSOI, mainly using Density Matrix Renormalization Group (DMRG) calculations
on wires with open boundary conditions. Since the Luttinger liquid
is the basic model to describe one-dimensional interacting electrons
also in the presence of spin-orbit interaction {[}\onlinecite{Starykh_08,LL_w_SOI_1,LL_w_SOI_7,LL_w_SOI_9,LL_w_SOI_12}{]}
we supplement our numerical analysis by a bosonization treatment of
the selected limiting cases under consideration. The main outcome
is the presence of long range spin current wave order in the ground
state of all of the insulating phases found in the system, together
with charge bond wave order.

The paper is organized as follows: in Section II we introduce the
Hamiltonian model and detail the order parameters of interest; in
particular we identify the presence of gapped phases within the approximation
of perturbatively interacting electrons, in the bosonization framework.
Then the setting to fully study electron-electron correlations \textendash{}
the Density Matrix Renormalization Group method \textendash{} is described
in Section III. The Section IV is devoted to present the numerical
results, with main focus on the most prominent gapped phase at half
filling and vanishing magnetization; analytical support is also briefly
discussed with details deferred to Appendices. Finally, in Sect. V
we summarize our results.

\section{Model and order parameters \label{sec:The-model}}

A microscopic Hamiltonian modeling a quantum wire with modulated RSOI
can be written in a tight-binding formulation as {[}\onlinecite{JJF_09}{]}
\begin{eqnarray}
H & = & -t\sum_{n,\alpha}\left(c_{n,\alpha}^{\dagger}c_{n+1,\alpha}^{\phantom{\dagger}}+H.c.\right)\nonumber \\
 & {\color{red}{\normalcolor +}} & i\sum_{n,\alpha,\beta}\gamma_{R}(n)\left(c_{n,\alpha}^{\dagger}\sigma_{\alpha\beta}^{y}c_{n+1,\beta}^{\phantom{\dagger}}-\mbox{H.c.}\right)\nonumber \\
 & - & \frac{h_{y}}{2}\sum_{n,\alpha,\beta}c_{n,\alpha}^{\dagger}\sigma_{\alpha\beta}^{y}c_{n,\beta}^{\phantom{\dagger}}-\mu\sum_{n,\alpha}c_{n,\alpha}^{\dagger}c_{n,\alpha}^{\phantom{\dagger}}\nonumber \\
 & + & U\sum_{n}\left(c_{n,\uparrow}^{\dagger}c_{n,\uparrow}^{\phantom{\dagger}}\right)\left(c_{n,\downarrow}^{\dagger}c_{n,\downarrow}^{\phantom{\dagger}}\right)\,\label{eq:Hamiltonian-z}
\end{eqnarray}
where $c_{n,\alpha}^{\dagger}$ ($c_{n,\alpha}^{\phantom{\dagger}}$)
are the creation (annihilation) operators for electrons on sites $n$
(numbered along the $\hat{x}$ axis) with spin ${\alpha}=\uparrow,\downarrow$
in the quantization axis $\hat{z}$ , $\vec{\sigma}$ are the Pauli
matrices, $t$ is the electron hopping amplitude, $\mu$ a chemical
potential, $h_{y}$ is a transverse external magnetic field along
$\hat{y}$ and $U$ is the strength of on-site Hubbard interaction.
We consider a modulated amplitude $\gamma_{R}(n)$ for the RSOI containing
a uniform term and an oscillating part with modulation length $\lambda=2\pi/Q$,
\begin{equation}
\gamma_{R}(n)=\gamma_{0}+\gamma_{1}\cos\left(Qn\right).\label{eq:modulation}
\end{equation}
In what follows, if not indicated specially, we take $U>0$ to describe
repulsive electron-electron interactions. As the one dimensional spin-momentum
Rashba coupling contains only $\sigma^{y}$ terms (defining the SO
axis), and we have restricted to magnetic fields along $\hat{y}$,
spin components are decoupled after a rotation of $\pi/2$ around
the $\hat{x}$ axis; in the following we indicate this spin polarization
by an index $\tau=\pm$ and the corresponding electron creation (annihilation)
operators by $d_{n,\tau}^{\dagger}$ ($d_{n,\tau}^{\phantom{\dagger}}$)
with 
\begin{equation}
\left(\begin{array}{c}
d_{n,+}\\
d_{n,-}
\end{array}\right)=e^{i\frac{\pi}{4}\sigma_{x}}\left(\begin{array}{c}
c_{n,\uparrow}\\
c_{n,\downarrow}
\end{array}\right).\label{spinor}
\end{equation}
In this basis the Hamiltonian in Eq. (\ref{eq:Hamiltonian-z}) reads
$H=H_{+}+H_{-}+H_{int}$, where 
\begin{eqnarray}
H_{\tau} & = & -t\sum_{n}\left(d_{n,\tau}^{\dagger}d_{n+1,\tau}^{\phantom{\dagger}}+H.c\right)\nonumber \\
 & + & i\tau\gamma_{0}\sum_{n}\left(d_{n,\tau}^{\dagger}d_{n+1,\tau}^{\phantom{\dagger}}-H.c.\right)\nonumber \\
 & + & i\tau\gamma_{1}\sum_{n}\!\cos(Qn)\left(d_{n,\tau}^{\dagger}d_{n+1,\tau}^{\phantom{\dagger}}-H.c.\right)\nonumber \\
 & - & \sum_{n}\left(\mu+\tau\frac{h_{y}}{2}\right)d_{n,\tau}^{\dagger}d_{n,\tau}^{\phantom{\dagger}}\label{eq:H_tau}
\end{eqnarray}
and 
\begin{eqnarray}
H_{int} & = & U\sum_{n}\,\left(d_{n,+}^{\dagger}d_{n,+}^{\phantom{\dagger}}\right)\left(d_{n,-}^{\dagger}d_{n,-}^{\phantom{\dagger}}\right)\,.\label{eq:Hamiltonian-U}
\end{eqnarray}
This can be also be written in terms of fermionic bilinears as
\begin{eqnarray}
H & = & -t\sum_{n,\tau}q_{n,n+1}^{\tau}+\sum_{n,\tau}\tau\gamma_{R}(n)j_{n,n+1}^{\tau}\label{eq:H-bilinears}\\
 &  & +\sum_{n,\tau}\left(\mu+\tau\frac{h_{y}}{2}\right)\rho_{n,\tau}+U\sum_{n}\rho_{n,+}\rho_{n,-}\,,\nonumber 
\end{eqnarray}
where we introduce on-site polarized densities as
\begin{equation}
\rho_{n}^{\tau}=d_{n,\tau}^{\dagger}d_{n,\tau},\label{eq: density operator}
\end{equation}
on-bond polarized densities as
\begin{equation}
q_{n,n+1}^{\tau}=d_{n,\tau}^{\dagger}d_{n+1,\tau}+d_{n+1\tau}^{\dagger}d_{n,\tau}\label{eq: q operator}
\end{equation}
and polarized current densities 
\begin{equation}
j_{n,n+1}^{\tau}=i\left(d_{n,\tau}^{\dagger}d_{n+1,\tau}-d_{n+1,\tau}^{\dagger}d_{n,\tau}\right).\label{eq: j operator}
\end{equation}

One of the aims of the present work is to describe charge and spin
wave orders in the insulating phases of the model in Eq. (\ref{eq:Hamiltonian-z}).
Thus we propose, for a wire with $L$ sites, the consideration of
the following ground state modulated averages of polarized densities:

\begin{equation}
\langle\rho^{\tau}\rangle_{Q}=\frac{1}{L}\sum_{n}\cos(Qn)\langle\rho_{n}^{\tau}\rangle,\label{eq:rho-tau}
\end{equation}

\begin{equation}
\langle q^{\tau}\rangle_{Q}=\frac{1}{L}\sum_{n}\cos(Qn)\langle q_{n,n+1}^{\tau}\rangle,\label{eq:q-tau}
\end{equation}

\begin{equation}
\langle j^{\tau}\rangle_{Q}=\frac{1}{L}\sum_{n}\cos(Qn)\langle j_{n,n+1}^{\tau}\rangle.\label{eq:j-tau}
\end{equation}
One can recover the corresponding charge densities by adding both
spin polarizations and the corresponding spin (magnetic) densities
along $\hat{y}$ by subtracting the different spin polarizations.
We then define the following order parameters for detecting the modulation
of charge and spin densities: 

- the on-site charge density wave
\begin{equation}
O_{CDW}=\langle\rho_{n}^{+}\rangle_{Q}+\langle\rho_{n}^{-}\rangle_{Q}\,,\label{eq:O_CDW}
\end{equation}

- the on-site spin density wave
\begin{equation}
O_{SDW}=\langle\rho_{n}^{+}\rangle_{Q}-\langle\rho_{n}^{-}\rangle_{Q}\,,\label{eq:O_SDW}
\end{equation}

- the charge bond order wave
\begin{equation}
O_{CBOW}=\langle q^{+}\rangle_{Q}+\langle q^{-}\rangle_{Q}\,,\label{eq:O_CBOW}
\end{equation}

- the spin bond order wave
\begin{equation}
O_{SBOW}=\langle q^{+}\rangle_{Q}-\langle q^{-}\rangle_{Q}\,,\label{eq:O_SBOW}
\end{equation}

- the charge current wave
\begin{equation}
O_{CCW}=\langle j^{+}\rangle_{Q}+\langle j^{-}\rangle_{Q}\,,\label{eq:O_CCW}
\end{equation}

- the spin current wave (a.k.a. chiral asymmetry current)
\begin{equation}
O_{SCW}=\langle j^{+}\rangle_{Q}-\langle j^{-}\rangle_{Q}\,.\label{eq:O_SCW}
\end{equation}

\begin{figure}
\begin{centering}
\includegraphics[scale=0.45]{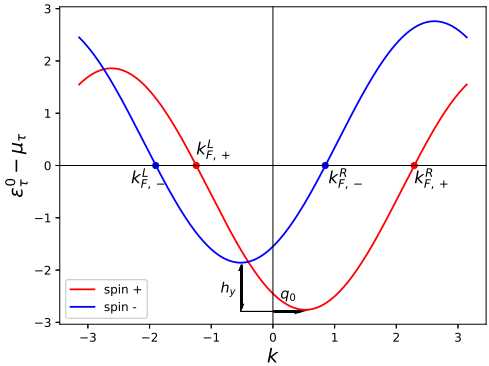} 
\par\end{centering}
\caption{An illustration of the single particle dispersion relations in the
presence of uniform Rashba SO interaction and transverse magnetic
field (arbitrary parameters). The horizontal shift $\pm q_{0}$ of
the bands is due to the uniform Rashba SO interaction, while the vertical
shift $h_{y}$ reflects the Zeeman splitting. A single horizontal
line shows the Fermi level for both bands. Four different Fermi momenta
are needed for the bosonization formalism.}
\label{fig:Fig2} 
\end{figure}

In order to exhibit basic properties, we first discuss the model in
absence of RSOI modulation ($\gamma_{1}=0$) and electron-electron
interactions ($U=0$). The Hamiltonians in Eq. (\ref{eq:H_tau}) can
then be trivially diagonalized in momentum space. One obtains 
\begin{equation}
H_{\tau}^{0}=\sum_{k=-\pi}^{\pi}\left(\epsilon_{\tau}^{0}(k)-\mu_{\tau}\right)\,d_{k,\tau}^{\dagger}d_{k,\tau\,}^{\phantom{\dagger}},\label{eq:H_tau_0}
\end{equation}
where $\epsilon_{\tau}^{0}(k)=-2\tilde{t}\cos(k-\tau q_{0})$ with
$\tilde{t}=\sqrt{t^{2}+\gamma_{0}^{2}}$ , $q_{0}=\arctan(\gamma_{0}/t)$,
and $\mu_{\tau}=\mu+\tau\frac{h_{y}}{2}$. As one can observe in Fig.
\ref{fig:Fig2}, plotted for generic parameters, the $\tau=\pm$ bands
are shifted horizontally by $\pm q_{0}$ because of the homogeneous
RSOI and vertically by $\mp h_{y}/2$ because of the external magnetic
field. The effective chemical potentials $\mu_{\tau}$ independently
control the filling fraction of each band, given by $\nu_{\tau}=\left(\nu+\tau m\right)/2$
in terms of the total electron filling fraction $\nu$ and the SO
axis magnetization fraction $m$.

Considering RSOI modulations ($\gamma_{1}\neq0$), still in absence
of electron-electron interactions ($U=0$), the Hamiltonians in Eq.
(\ref{eq:H_tau}) are quadratic and can be exactly diagonalized, even
analytic results can be obtained for short length modulations\textcolor{blue}{{}
}(see Appendix A); this provides most clear results that can be obtained
by exact diagonalization. However, for analytical discussions we find
it convenient to treat the RSOI modulations as perturbations. This
allows us to determine, within the bosonization approach {[}\onlinecite{GNT_book}{]},
the commensurate values of the band fillings $\nu_{\tau}$ at which
the RSOI modulation opens band gaps and leads the electron system
into a \emph{band insulator} phase. 

\subsection*{Bosonization picture}

The advantage of the bosonization procedure relies on its prediction
power and the universal extent of its results, though it is well suited
for the weak-coupling limit; here we assume $|U|,|\gamma_{1}|\ll\tilde{t}$
and treat both RSOI modulations and electron-electron interactions
on equal footing as perturbations with respect to free Hamiltonians
in Eq. (\ref{eq:H_tau_0}). The bosonization formalism may look slightly
different from usual presentations, as we apply it to shifted bands:
it is necessary to identify four Fermi points (see Fig. \ref{fig:Fig2})
\begin{eqnarray}
k_{F,\tau}^{R} & = & \tau q_{0}+k_{F,\tau}^{0}\,,\label{eq:k_F_Right}\\
k_{F,\tau}^{L} & = & \tau q_{0}-k_{F,\tau}^{0}\,,\label{eq:k_F_Left}
\end{eqnarray}
where $k_{F,\tau}^{0}=\nu_{\tau}\pi$ are the usual Fermi momenta
in absence of Rashba interactions, at band filling $\nu_{\tau}$.
Then the procedure is straightforward. Using the standard recipes
(see for instance {[}\onlinecite{GNT_book}{]}) on the free Hamiltonians
and the perturbations one obtains the following bosonized Hamiltonian
\begin{widetext} 
\begin{eqnarray}
H_{bos} & = & \sum_{\tau}\int dx\Big\{\,\frac{v_{F}^{\tau}}{2}\left[\left(\partial_{x}\varphi_{\tau}\right)^{2}+\left(\partial_{x}\vartheta_{\tau}\right)^{2}\right]+\frac{2\gamma_{0}\gamma_{1}}{\pi\alpha_{0}\tilde{t}}\,\sum_{j=\pm}\sin\left[\left(jQ+2k_{F,\tau}^{0}\right)x+k_{F,\tau}^{0}+\sqrt{4\pi}\varphi_{\tau}(x)\right]\Big\}\nonumber \\
 & + & \frac{U}{\pi}\int dx\left[\left(\partial_{x}\varphi_{+}\right)\left(\partial_{x}\varphi_{-}\right)+\frac{1}{\pi\alpha_{0}^{2}}\sin\left(\sqrt{4\pi}\varphi_{+}(x)+2k_{F,+}^{0}x\right)\sin\left(\sqrt{4\pi}\varphi_{-}(x)+2k_{F,-}^{0}x\right)\right]\label{eq:H-bos}
\end{eqnarray}
\end{widetext} where $\varphi_{\tau}(x)$ and $\vartheta_{\tau}(x)$
are dual bosonic fields, $v_{F}^{\tau}=2\tilde{t}\,\sin\left(k_{F,\tau}^{0}\right)$
are their Fermi velocities and $\alpha_{0}$ is a cutoff required
to be of the order of the lattice constant. Notice that $v_{F}^{+}\neq v_{F}^{-}$
as soon as the system is magnetized ($m\neq0$). 

In absence of Hubbard interactions one can see that the effect of
perturbations introduced by the modulated RSOI are present in the
continuum limit\emph{ }only provided that $\gamma_{0}\neq0$ \emph{and
}$\gamma_{1}\neq0$ , and survive \emph{only at commensurate band-fillings}
given by separate different conditions 
\begin{equation}
Q\pm2k_{F,\tau}^{0}\cong0\,\left(\text{mod }2\pi\right)\,\label{eq:condition.1}
\end{equation}
for each spin polarization band. When one of them is met, a relevant
perturbation opens a gap to the corresponding spin polarized excitations.
The case where the commensurability holds for just one of the spin
polarizations corresponds to the half-metallic phases considered in
{[}\onlinecite{Cabra_etal_17}{]}. 

In the present work we focus on fully gapped phases, met when $Q\pm2k_{F,+}^{0}\cong0$
\emph{and} $Q\pm2k_{F,-}^{0}\cong0$. This requires at least $m=0$
or $\nu=1$, conditions that may be met by varying the magnetic field
and the chemical potential. The qualitatively different attainable
gapped phases are then: 

- non magnetized insulator at half-filling ($m=0$, $\nu=1$).

- non magnetized insulator away from half-filling ($m=0$, $\nu\neq1$).

- magnetized insulator at half-filling ($m\neq0$, $\nu=1$).

On the other hand, the Hubbard interaction term in Eq. (\ref{eq:H-bos})
couples the $\varphi_{+}(x)$ and $\varphi_{-}(x)$ fields and does
not allow for straightforward inspection. We will comment on its perturbative
effect on the different gapped phases after presenting our numerical
results for interacting electrons.

For later reference, we recall that the bilinear operators in Eqs.
(\ref{eq: density operator}-\ref{eq: j operator}) take the following
bosonized forms
\begin{equation}
\rho_{n}^{\tau}\simeq\frac{1}{\sqrt{\pi}}\partial_{x}\varphi_{\tau}(x)+\frac{1}{\pi\alpha}\sin\left(\sqrt{4\pi}\varphi_{\tau}+2k_{F,\tau}^{0}x\right)\,,\label{eq: rho-op-bos}
\end{equation}

\begin{widetext} 
\begin{equation}
q_{n,n+1}^{\tau}\simeq\frac{2\cos(q_{0})}{\sqrt{\pi}}\cos\left(k_{F,\tau}^{0}\right)\partial_{x}\varphi_{\tau}(x)-\frac{2\tau\sin(q_{0})}{\sqrt{\pi}}\sin\left(k_{F,\tau}^{0}\right)\partial_{x}\vartheta_{\tau}(x)+\frac{2\cos(q_{0})}{\pi\alpha}\sin\left(\sqrt{4\pi}\varphi_{\tau}+2k_{F,\tau}^{0}x+k_{F,\tau}^{0}\right)\,,\label{eq: q-op-bos}
\end{equation}
and
\begin{equation}
j_{n,n+1}^{\tau}\simeq-\frac{2\cos(q_{0})\sin(k_{F,\tau}^{0})}{\sqrt{\pi}}\partial_{x}\vartheta_{\tau}(x)-\frac{2\tau\sin(q_{0}\tau)\cos(k_{F,\tau}^{0})}{\sqrt{\pi}}\partial_{x}\varphi_{\tau}(x)-\frac{2\tau\sin(q_{0})}{\pi\alpha}\sin\left(\sqrt{4\pi}\varphi_{\tau}+2k_{F,\tau}^{0}x+k_{F,\tau}^{0}\right)\,.\label{eq: Rashba-bosonized}
\end{equation}

\end{widetext} These will allow for a semiclassical inspection of
the order parameters in Eqs. (\ref{eq:O_CDW}-\ref{eq:O_SCW}) in
the different gapped phases.

\section{DMRG investigation of the effect of electron-electron interactions\label{sec:DMRG}}

In order to investigate the\textcolor{blue}{{} }non-perturbative effects
of electron-electron interactions in the present model we have performed
extensive numerical computations in the Density Matrix Renormalization
Group (DMRG) framework {[}\onlinecite{White_1992}{]}, with repulsive
Hubbard couplings ranging from $U=0$ up to $U=25\,t$, and additional
explorations with attractive interactions $U<0$. Our results provide
a description of the charge and spin gaps, and correlation induced
effects in the alternating charge and spin order structures.

In this work we employ the finite-size DMRG algorithm, as implemented
in the ALPS library {[}\onlinecite{Bauer_2011}{]}. We have run simulations
for systems up to $L=128$ sites, using open boundary conditions (OBC).
We have computed the lowest energy states in eigenspaces of spin polarized
number operators ${\hat{N}}_{\tau}=\sum_{n}\rho_{n}^{\tau}$, in order
to estimate the charge and spin excitation gaps. We have also computed
local expectation values and nearest neighbors correlations to estimate
the order parameters.

The choice of boundary conditions deserves some observations. On the
one side, the use of periodic boundary conditions (PBC) requires a
careful commensurability of system lengths to avoid a net magnetic
flux associated to the accumulation of the complex phases of $t\pm i\gamma_{R}(n)$
in Eq. (\ref{eq:H_tau}). On the other side, an OBC chain with spatially
modulated hopping acquires a topological character {[}\onlinecite{SSH_1979,Shen_2012}{]}
that may introduce edge bound states with energies laying inside the
gaps we aim to compute, depending on the modulation phase chosen for
the left-most bond in the wire and the commensurability between the
wire and modulation lengths. We have taken rational modulations $Q=2\pi r/p$
and chain lengths which are integer multiples of $p$, setting $\cos(nQ)=+1$
for the left-most bond; this renders the open chain in the topologically
trivial sector, avoiding (here) undesired gapless edge states. Following
this recipe we have analyzed chains of $L=48$, $64$, $96$ and $128$
sites for modulations with wave number $Q=\pi$ and $L=48$, $66$,
$96$ and $126$ sites for $Q=2\pi/3$. Data points have been obtained
keeping $600$ states during $20$ sweeps. The estimated error for
energy gaps is less than $10^{-5}t$, which ensures enough energy
precision for the results we report. Two-point correlations are computed
within an error of $10^{-6}$.

The Hamiltonian in Eq.~(\ref{eq:Hamiltonian-z}) commutes with the
total charge operator ${\hat{Q}}={\hat{N}}_{+}+{\hat{N}}_{-}$ and
the total spin $y$-component operator ${\hat{M}}=\frac{1}{2}\left({\hat{N}}_{+}+{\hat{N}}_{-}\right)$.
In consequence, the eigenvalues of ${\hat{N}}_{+}$ and ${\hat{N}}_{-}$
are good quantum numbers describing the occupation of states with
given spin polarization $\tau=\pm$. For a system with $L$ sites
and $N_{\tau}$ occupied states in each spin sector, the filling fraction
$\nu$ and the transverse magnetization density $m$ mentioned in
the previous Section are determined as 
\begin{eqnarray}
\nu & = & (N_{+}+N_{-})/L\,,\\
m & = & (N_{+}-N_{-})/L\,.\label{eq:nu-m-1}
\end{eqnarray}
Given $\nu$ and $m$, defining a band insulating phase, we determine
by DMRG the lowest energy state (without external field and chemical
potential) in the subspace with $N_{\tau}=L(\nu+\tau\,m)/2$ occupied
states with spin $\tau$; we denote by $E_{0}(N_{+},N_{-})$ the corresponding
energy eigenvalue. Chemical potential and external magnetic field
energy contributions are proportional to the total charge and spin
$y$-component respectively, so they just produce an energy shift
that can be added later when needed. Expectation values of local operators
and correlations are computed in such states.

For describing charge and spin excitations we consider the standard
\emph{two-particle} excitation gaps. The charge gap is defined as
the average energy cost of adding or removing two particles with different
spin orientation, thus without change in the magnetization,\begin{widetext}
\begin{equation}
\Delta_{c}=\frac{E_{0}(N_{+}+1,N_{-}+1)+E_{0}(N_{+}-1,N_{-}-1)-2E_{0}(N_{+},N_{-})}{2}\,.\label{eq:deltaC}
\end{equation}
Similarly, the spin gap is defined as the average energy cost of adding
a particle with a given spin orientation and removing another with
the opposite, without changing the total charge, 
\begin{equation}
\Delta_{s}={\color{red}}\frac{E_{0}(N_{+}+1,N_{-}-1)+E_{0}(N_{+}-1,N_{-}+1)-2E_{0}(N_{+},N_{-})}{2}\,.\label{eq:deltaS}
\end{equation}
\end{widetext}Defining also the \emph{one-particle} gaps as the average
energy cost of adding or removing a particle with a given spin polarization,
\begin{equation}
\Delta_{+}=\frac{E_{0}(N_{+}+1,N_{-})+E_{0}(N_{+}-1,N_{-})-2E_{0}(N_{+},N_{-})}{2}\label{eq:delta+}
\end{equation}
and 
\begin{equation}
\Delta_{-}=\frac{E_{0}(N_{+},N_{-}+1)+E_{0}(N_{+},N_{-}-1)-2E_{0}(N_{+},N_{-})}{2},\label{eq:delta-}
\end{equation}
for non-interacting electrons the two-particle gaps are simply related
to the highest occupied and lowest unoccupied one-particle energies
by $\Delta_{c}=\Delta_{s}=\Delta_{+}+\Delta_{-}$. The presence of
electron interactions generally changes these relations; the more
different charge and spin gaps are, the more correlated the system
is.

\section{Results\label{sec: Results}}

In this Section we focus on the three situations pointed out at the
end of Section \ref{sec:The-model}, where bosonization anticipates
insulator phases. We choose specific RSOI modulation lengths $Q$,
electron fillings $\nu$ and magnetizations $m$ in order to investigate
the effects of the electron-electron interactions on the charge and
spin gaps, and on the corresponding wave order patterns, in the three
selected situations. Of course, the opening of band gaps in absence
of electron interactions is easily verified by Fourier diagonalization
of $H_{\tau}$ in Eq. (\ref{eq:H_tau}). For numerical computations
we set in the following the hopping amplitude $t=1$, the homogeneous
Rashba coefficient $\gamma_{0}\approx0.577$ (exactly $\gamma_{0}=\tan\left(q_{0}\right)$
with $q_{0}=\pi/6$) and the amplitude of the Rashba coefficient oscillation
$\gamma_{1}=0.2$; we found no qualitative differences for RSOI parameters
in the range $0<\gamma_{0},\gamma_{1}<1$. 

The most salient feature in our results is the presence of a long
range spin current wave order in the ground state of all of the insulating
phases found in the system, together with charge bond order waves. 

\subsection{Non magnetized insulator at half-filling\label{sec:Non-magnetized-insulator-half-filling}}

We start discussing in detail the gaps and the order parameters for
the most prominent insulating phase of the Hamiltonian in Eq. (\ref{eq:Hamiltonian-z}),
that with half-filling $\nu=1$ and no magnetization $m=0$. One finds
that $k_{F,\tau}^{0}=\pi/2$ and $v_{F,\tau}\equiv v_{F}$ irrespective
of the spin projection. In order to fulfill the commensurability conditions
in Eq. (\ref{eq:condition.1}) the RSOI modulations are required to
have wave number $Q=\pi$, that is a two-site wave length. Such a
short length modulation could be observed in a layered material, by
designing a quantum wire on top of an anti-ferroelectric substrate
{[}\onlinecite{Streltsov-20015}{]}. 

\begin{figure}
\begin{centering}
\includegraphics[scale=0.45]{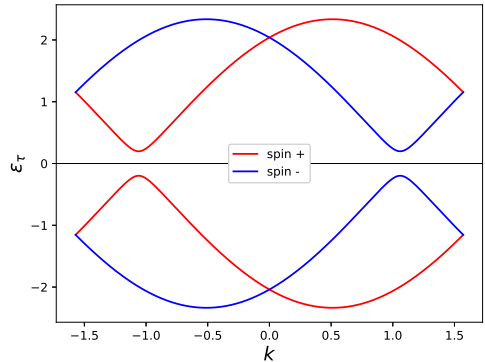} 
\par\end{centering}
\caption{One particle energy dispersion bands at half-filling, no magnetization
and RSOI modulation with wave number $Q=\pi$, for a wire of $L=1000$
sites and periodic boundary conditions. Here and in remaining figures
we set $t=1$, $\gamma_{0}\approx0.577$ and $\gamma_{1}=0.2$.}
\label{fig: 2 - A-dispersion} 
\end{figure}

Given the short length modulation, it is worth to review the analytical
description of the band structure. In absence of interactions the
one particle spectrum has two bands with dispersion relations $\pm\epsilon_{\tau}(k)$,
where
\begin{equation}
\epsilon_{\tau}(k)=2\,\sqrt{\tilde{t}^{2}\cos^{2}(k-\tau q_{0})+\gamma_{1}^{2}\cos^{2}(k)}\,,\label{eq:A-dispersion}
\end{equation}
and $-\frac{\pi}{2}\leq k<\frac{\pi}{2}$ (see Appendix A). Only at
finite $\gamma_{0}$ \emph{and} $\gamma_{1}$, in agreement with bosonization
prediction in Section \ref{sec:The-model}, these bands are separated
by an energy gap 
\begin{equation}
\Delta=2\,\sqrt{2t'^{2}-2\sqrt{t'^{4}-4\gamma_{0}^{2}\gamma_{1}^{2}}}\label{eq: A-analytic gap}
\end{equation}
found at incommensurate momentum 
\begin{equation}
k^{*}=\tau\,\text{arccot}\left(\frac{2t\gamma_{0}}{t^{2}-\gamma_{0}^{2}+\gamma_{1}^{2}}\right)\,.\label{eq: A-gap-momentum}
\end{equation}
The dispersion bands are shown in Fig. \ref{fig: 2 - A-dispersion},
as obtained numerically under PBC. 

At half-filling, $U=0$, and zero temperature the lower bands are
completely occupied; with this information one can compute ground
state expectation values. From Eqs. (\ref{eq:O_CDW-ready}, \ref{eq:O_SDW-ready},
\ref{eq:O_bond-ready}) in Appendix A\textcolor{blue}{{} }we learn that
in the present phase there is no site density wave order
\begin{equation}
O_{CDW}=O_{SDW}=0
\end{equation}
nor bond spin wave order nor charge current wave order
\begin{equation}
O_{SBOW}=O_{CCW}=0\,.
\end{equation}
In contrast,
\begin{equation}
O_{CBOW}=-\frac{1}{\pi}\sum_{\tau}\int_{-\pi/2}^{\pi/2}\frac{\tau\gamma_{1}\sin(2k)}{\epsilon_{\tau}(k)}\,dk\neq0\label{eq:O_bond-CDW-integral}
\end{equation}
and 
\begin{equation}
O_{SCW}=\frac{1}{\pi}\sum_{\tau}\tau\int_{-\pi/2}^{\pi/2}\frac{2\tau\gamma_{1}\cos^{2}(k)}{\epsilon_{\tau}(k)}\,dk\neq0\label{eq:O_SCW-integral}
\end{equation}
for $\gamma_{0}\neq0$ and $\gamma_{1}\neq0$.

\begin{figure}
\begin{centering}
\includegraphics[scale=0.5]{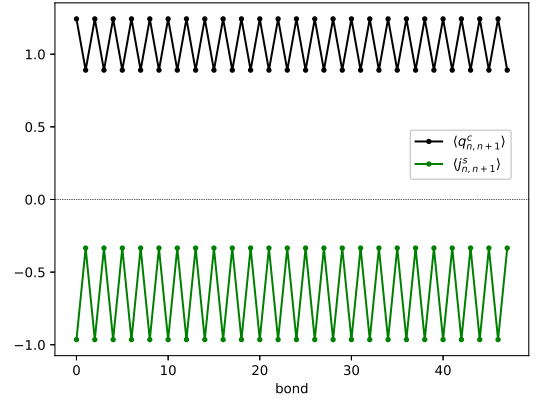} 
\par\end{centering}
\caption{Local oscillation of bond charge density $\langle q_{n,n+1}^{c}\rangle$
and spin current density $\langle j_{n,n+1}^{s}\rangle$ in the ground
state of a wire of $L=48$ sites (PBC), at half-filling and no magnetization.
Charge density is homogeneous, while spin density, on-bond spin and
charge current densities vanish (not shown). }
\label{fig: A-local qc js} 
\end{figure}

The presence of alternating long range order in the spin current,
expressed by $O_{SCW}\neq0$, is a distinguished feature of the present
model. It might be better appreciated from the spatial expectation
value profile of the operators in Eqs. (\ref{eq: density operator}-\ref{eq: j operator}),
easily computed in absence of electron interactions. One finds that
the local occupation number is homogeneous with $\langle\rho_{n}^{\tau}\rangle=0.5$
for both spin polarizations, so that there is neither CDW nor SDW
order. In contrast, $\langle q_{n,n+1}^{\tau}\rangle$ oscillates
with period two and the \emph{same values} for both polarizations,
while $\langle j_{n,n+1}^{\tau}\rangle$ oscillates with the same
period but \emph{opposite values} for different polarizations. Then
the corresponding modulations in $\langle q_{n,n+1}^{S}\rangle=\langle q_{n,n+1}^{+}-q_{n,n+1}^{-}\rangle$
and $\langle j_{n,n+1}^{C}\rangle=\langle j_{n,n+1}^{+}+j_{n,n+1}^{-}\rangle$
cancel out, while $\langle q_{n,n+1}^{C}\rangle=\langle q_{n,n+1}^{+}+q_{n,n+1}^{-}\rangle$
and $\langle j_{n,n+1}^{S}\rangle=\langle j_{n,n+1}^{+}-j_{n,n+1}^{-}\rangle$
add up, as shown in Fig. \ref{fig: A-local qc js}. This explains
the reason why $O_{SBOW}=O_{CCW}=0$ but $O_{CBOW}$ and $O_{SCW}$
do not vanish. 

Indeed, there is an underlying reason for the observed relations between
the ground state expectation values of bond densities and current
densities: under the time evolution governed by the Hamiltonian in
Eq. (\ref{eq:Hamiltonian-z}) the currents 
\begin{equation}
J_{n\to n+1}^{\tau}=t\,j_{n,n+1}^{\tau}+\tau\gamma_{R}(n)\,q_{n,n+1}^{\tau}\label{eq: conserved current}
\end{equation}
are conserved, whether with or without electron interactions (see
Appendix B). That is, the usual expression for particle density currents
$j_{n,n+1}^{\tau}$ is modified by the presence of the RSOI. This
fact, together with inversion symmetry (w.r.t. bond centered inversion
points) shows that in stationary states $\langle J_{n\to n+1}^{(\tau)}\rangle=0$
at any bond. Then bond densities and currents are deeply connected
by
\begin{equation}
t\,\langle j_{n,n+1}^{\tau}\rangle=-\tau\gamma_{R}(n)\,\langle q_{n,n+1}^{\tau}\rangle\,,\label{eq: q-j-relation}
\end{equation}
as has been verified in numerical data all along the present work. 

\medskip{}

\begin{figure}
\begin{centering}
\includegraphics[scale=0.5]{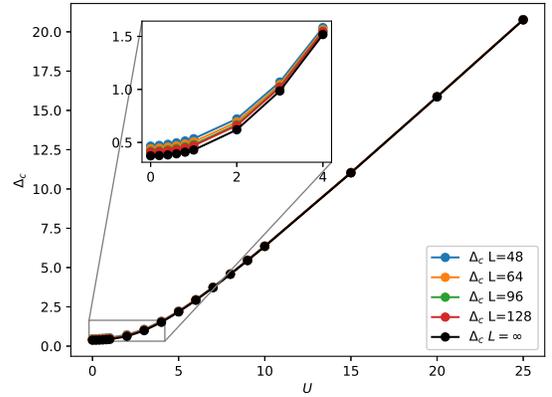}
\par\end{centering}
\caption{Evolution of the charge gap with $U$, at half-filling, no magnetization
and $Q=\pi$. For large $U$ the gap grows linearly, indicating a
Mott insulator phase. Several wire lengths and the infinite size extrapolation
are shown. \label{fig:charge-gap-caseA}}
\end{figure}

\begin{figure}
\begin{centering}
\includegraphics[scale=0.5]{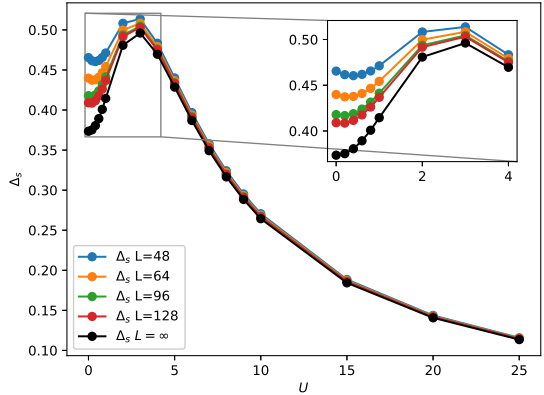}
\par\end{centering}
\caption{Evolution of the spin gap with $U$, at half-filling, no magnetization
and $Q=\pi$. After reaching a maximum the spin gap decays with increasing
$U$\textcolor{blue}{.} It remains finite, suggesting an asymptotic
approach to zero for $U\to\infty$. Several wire lengths and extrapolation
are shown. \label{fig:spin-gap-caseA} }
\end{figure}

The effects of electron-electron interactions in the previous picture
is the main purpose of the present work. We have be explored these
effects numerically. Extensive DMRG computations (see Section \ref{sec:DMRG}
for details) show that the charge and spin gaps, which coincide at
$U=0$, \emph{do not close at any finite} $U$ but behave differently
suggesting a crossover from the band insulator to a correlated Mott
insulator regime. Under repulsive interactions $U>0$ the charge gap
grows, getting asymptotically linear for large $U$ as shown in Fig.
\ref{fig:charge-gap-caseA}. Instead the spin gap reaches a maximum
slightly above its band value and then decreases, as shown in Fig.
\ref{fig:spin-gap-caseA}. This suggests that the spin gap remains
finite for any finite $U$ and asymptotically approaches zero for
$U\to\infty$. 

\begin{figure}
\begin{centering}
\includegraphics[scale=0.5]{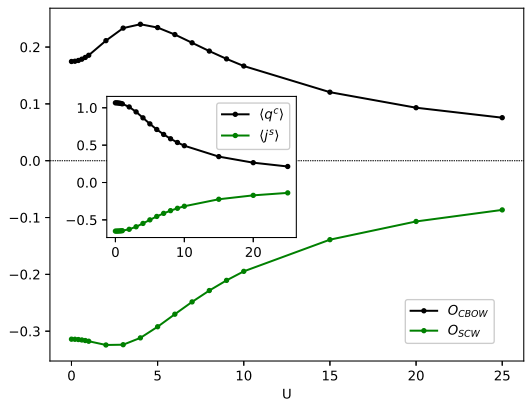}
\par\end{centering}
\caption{Charge bond wave order and spin current order parameters evolution
under electron-electron interactions, for a wire of $L=128$ sites.
The inset shows the plain averages of bond charge density $\langle q^{C}\rangle$
and spin current density $\langle j^{S}\rangle$. \label{fig:A- C-BOW SCW}}
\end{figure}

The order parameters are computed from nearest neighbors correlation
functions\textcolor{blue}{. }We have found that the long range alternating
order signaled by non vanishing $O_{CBOW}$ and $O_{SCW}$ is \emph{robust
against electron-electron interactions}, as shown in Fig. \ref{fig:A- C-BOW SCW}.
Also charge density remains homogeneous at one particle per site,
as well as spin density, spin bond order and charge current order
remain null (not shown). 

\begin{figure}
\begin{centering}
\includegraphics[scale=0.5]{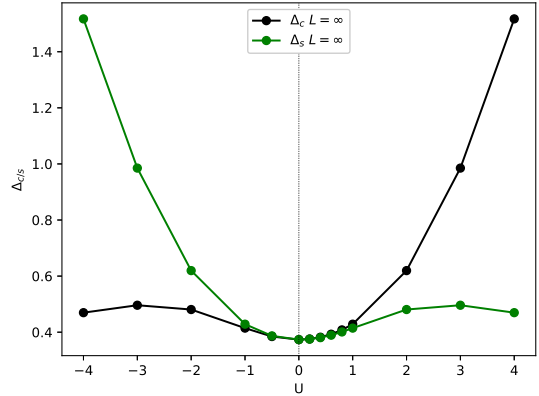}
\par\end{centering}
\caption{Charge and spin gaps are interchanged when the Hubbard interaction
is attractive instead of repulsive. We show here both gaps together,
which also allows for comparison of Figs. \ref{fig:charge-gap-caseA}
and \ref{fig:spin-gap-caseA}. Extrapolated in $1/L$ from $L=48,\,64,\,96,\,128$
sites. $t=1$, $\gamma_{0}=\tan\left(\pi/6\right)$, $\gamma_{1}=0.2$
and $Q=\pi$. \label{fig:Charge-and-spin-gaps}}
\end{figure}

Interestingly, we have found clear signals of a spin-charge duality
in the behavior of charge and spin gaps. A\textcolor{blue}{s} shown
in Fig. \ref{fig:Charge-and-spin-gaps}, they are interchanged when
comparing the repulsive regime $U>0$ with the attractive regime $U<0$,
a similar behavior to that of the Hubbard model at half-filling {[}\onlinecite{Yang-1990-1991}{]}.
This duality is not immediately expected, as the RSOI explicitly breaks
the $SU(2)$ symmetry of the Hamiltonian in Eq. (\ref{eq:Hamiltonian-z}).
However, a closer look shows that the present system can indeed be
mapped onto a spatially modulated hopping Hubbard model, without RSOI,
by means of an invertible $SU(2)$ gauge transformation (see Appendix
C). As the mapping preserves the charge and spin quantum numbers,
this implies that, in our model, all of the charge (spin) observables
in the repulsive regime are dual to the spin (charge) observables
in the attractive regime. Also the $SU(2)\times SU(2)$ spin and charge
symmetry, another important property of Hubbard models at half-filling
{[}\onlinecite{Yang-1990-1991}{]}, is present in our model. We point
out that the gauge mapping getting rid of the RSOI provides theoretical
insight into the original model at the price of introducing a twist
in the boundary conditions of finite length wires. This fact reflects
itself through intricacies in the analysis of edge effects in open
chains {[}\onlinecite{Goth_2014}{]}. We recall that our DMRG procedure,
dealing directly with the Hamiltonian in Eqs. (\ref{eq:H_tau},\ref{eq:Hamiltonian-U}),
has been carefully tuned to avoid undesired edge states. Analogously,
in finite periodic wires the mapping introduces a net flux which is
sensitive to the system length and makes unstable the infinite size
extrapolation.

\medskip{}

The present numerical results find support within the bosonization
analysis, at least at a perturbative level. At half-filling and zero
magnetization the Fermi momenta in absence of RSOI are $k_{F,+}^{0}=k_{F,-}^{0}=\pi/2$
and the sinusoidal term in the second line in Eq. (\ref{eq:H-bos})
is commensurate with the lattice spacing. For the same reason the
Fermi velocities of $\tau=\pm$ excitations in the first line are
the same, then the quadratic term introduced by the Hubbard interaction
can be incorporated into the free (Gaussian) Hamiltonian by the introduction
of the usual charge and spin bosonic fields {[}\onlinecite{GNT_book}{]}.
Following standard steps we find that at a semiclassical level there
is no competition between the remaining perturbative terms. The order
structure can then be inspected by evaluation of Eqs. (\ref{eq: rho-op-bos}-\ref{eq: Rashba-bosonized})
in the field configurations minimizing the semiclassical potential;
the results support the presence of long range $O_{CBOW}$ and $O_{SCW}$
order. The lack of competition between perturbative terms also supports
the absence of order transitions driven by the repulsive Hubbard interaction.
Finally, the linear growth of the charge gap is related to the commensurability
of the so-called Umklapp term {[}\onlinecite{GNT_book}{]} with wave
number $2\left(k_{F,+}^{0}+k_{F,-}^{0}\right)=2\pi$.

We also notice that the bosonization of the present system with RSOI
is related, through the gauge mapping discussed in Appendix C, to
the bosonization of one dimensional Hubbard systems with $SU(2)$
invariant perturbations. Along this line we have checked that our
results are consistent with the vast literature written about those
systems, and the competence amongst different relevant perturbations,
in the context of the Peierls-Hubbard model {[}\onlinecite{PH_0,EHM_1,PH_1,PH_2,PH_3,PH_4,PH_5,PH_6,PH_7,PH_8,PH_9,PH_10}{]}. 

\subsection{Magnetized insulator at half filling}

For completeness, we briefly report the results obtained in the other
gapped phases.

We showed in Section \ref{sec:The-model} that the existence of a
gapped phase with net magnetization requires the filling to be fixed
to one electron per site ($\nu=1$). As a representative case we analyze
here a system with RSOI modulations of wave number $Q=2\pi/3$ (three
sites wave length) and an external magnetic field along the $\hat{y}$
axis setting a net magnetization $m=1/3$, so that the commensurability
conditions in Eq. (\ref{eq:condition.1}) are satisfied with $k_{F,+}^{0}=2\pi/3$
and $k_{F,-}^{0}=\pi/3$. For numerical computations we set $t=1$,
$\gamma_{0}=t\,\tan(\pi/6)$ and $\gamma_{1}=0.2$, the same parameters
as in Section \ref{sec:Non-magnetized-insulator-half-filling}.

\begin{figure}
\begin{centering}
\includegraphics[scale=0.45]{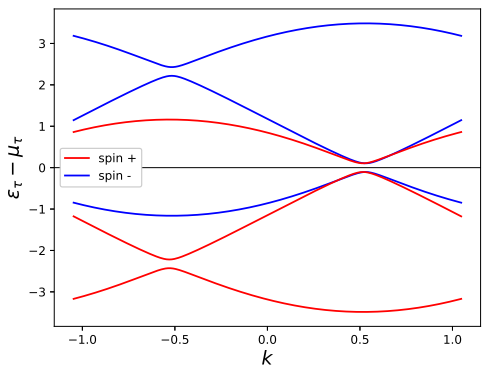} 
\par\end{centering}
\caption{Dispersion relations for elementary excitations in a half-filled magnetized
insulator, with $Q=2\pi/3$ and $\nu=1$. An appropriate magnetic
field $h_{y}$ sets a net magnetization $m=1/3$.}
\label{fig: half filling m not 0 spectrum}
\end{figure}

\begin{figure}
\begin{centering}
\includegraphics[scale=0.5]{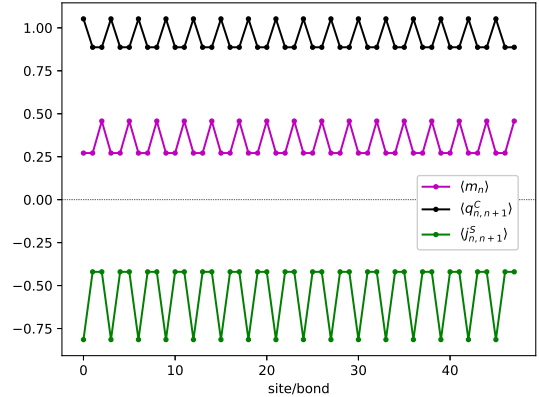} 
\par\end{centering}
\caption{Local oscillation of local magnetization $\langle m_{n}\rangle$,
bond charge density $\langle q_{n,n+1}^{C}\rangle$ and spin current
density $\langle j_{n,n+1}^{S}\rangle$ in a wire of $L=48$ sites
(PBC), with $Q=2\pi/3$, $\nu=1$ and $m=1/3$. Charge density is
homogeneous $\langle\rho_{n}^{C}=1\rangle$, while bond spin and charge
current densities vanish (not shown).}
\label{fig: half filling m not 0 band order-1}
\end{figure}

Disregarding electron interactions, the band structure is shown in
Fig. \ref{fig: half filling m not 0 spectrum}. This system shows
period three oscillations in the ground state expectation values of
the site magnetization $m_{n}=\rho_{n}^{+}-\rho_{n}^{-}$, in the
bond charge density $q_{n,n+1}^{C}=q_{n,n+1}^{+}+q_{n,n+1}^{-}$ and
in the spin current density $j_{n,n+1}^{S}=j_{n,n+1}^{+}-j_{n,n+1}^{-}$
as shown in Fig. \ref{fig: half filling m not 0 band order-1}. 

\begin{figure}
\begin{centering}
\includegraphics[scale=0.5]{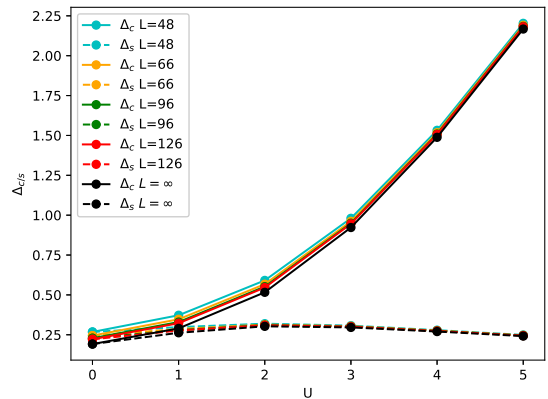} 
\par\end{centering}
\caption{Evolution of the charge and spin gaps with $U$, at half filling,
magnetization $m=1/3$ and RSOI modulations with $Q=2\pi/3$. For
large $U$ the charge gap grows linearly, indicating a Mott insulator
phase. There is no spin-charge duality for $U<0$ (not shown).}
\label{fig: half filling m not 0 gaps}
\end{figure}

In the presence of repulsive interactions the charge gap and spin
gaps evolve in a similar way as they do in the half-filled non-magnetized
phase, as shown in Fig. \ref{fig: half filling m not 0 gaps}. This
is supported by the bosonization analysis: both the oscillatory interacting
terms in Eq. (\ref{eq:H-bos}) and the Umklapp term are commensurate
with the lattice spacing. However, the net magnetization breaks the
spin-charge duality and the gaps for attractive $U<0$ are not symmetric
with respect to the repulsive regime.

\begin{figure}
\begin{centering}
\includegraphics[scale=0.5]{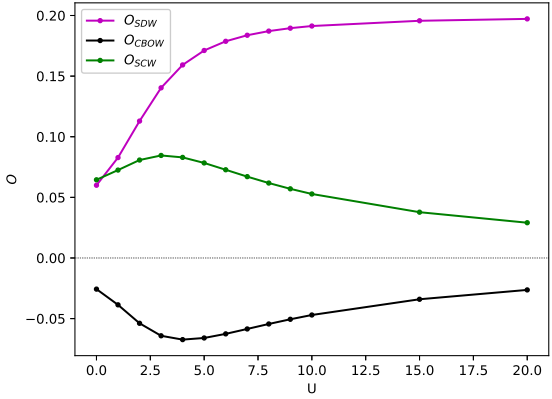} 
\par\end{centering}
\caption{Evolution of the non vanishing order parameters with $U$, for a wire
of $L=126$ sites at half filling $\nu=1$, net magnetization $m=1/3$
and RSOI modulations with $Q=2\pi/3$.}
\label{fig: half filling m not 0 orders}
\end{figure}

The order parameters defined in Eqs. (\ref{eq:O_CDW}-\ref{eq:O_SCW}),
with $Q=2\pi/3$, keep track of the oscillations of $\langle m_{n}\rangle$,
$\langle q_{n,n+1}^{C}\rangle$ and $\langle j_{n,n+1}^{S}\rangle$.
One can see in Fig. \ref{fig: half filling m not 0 orders} that they
are robust under the electron interactions, with a tendency to stabilize
magnetization oscillations and fade out bond density and current oscillations.
This appears to be consistent with the $U\to+\infty$ limit where
the system, at half-filling, is driven onto an anisotropic spin 1/2
Heisenberg model with modulated DMI {[}\onlinecite{Moriya-1960}{]}
at $m=1/3$ magnetization. Moreover, the limiting value of $O_{SDW}$
suggests the formation of an ordered quantum ground state alternating
two-site spin singlets and isolated spin + sites {[}\onlinecite{Hida-Affleck-2005}{]}. 

\subsection{Non magnetized insulator away from half-filling}

A gapped phase with electron filling away from one particle per site
requires that the net magnetization vanishes ($m=0$). A representative
case is chosen here as a system with RSOI modulations of wave number
$Q=2\pi/3$ (three sites wave length) with a chemical potential $\mu$
setting the electron filling at $\nu=2/3$, satisfying the commensurability
conditions in Eq. (\ref{eq:condition.1}) with $k_{F,+}^{0}=k_{F,-}^{0}=\pi/3$.
As before, for numerical computations we set $t=1$, $\gamma_{0}=t\,\tan(\pi/6)$
and $\gamma_{1}=0.2$.

\begin{figure}
\begin{centering}
\includegraphics[scale=0.45]{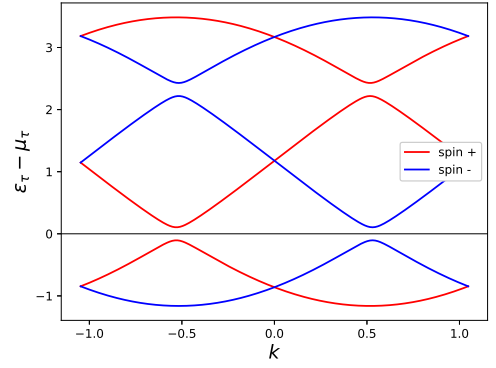} 
\par\end{centering}
\caption{Dispersion relations for elementary excitations in the non half-filled
non magnetized insulator, $m=0$ and $Q=2\pi/3$. An appropriate chemical
potential $\mu$ sets the filling fraction at $\nu=2/3$. }
\label{fig: third filling m=00003D0 spectrum}
\end{figure}

\begin{figure}
\begin{centering}
\textcolor{red}{\includegraphics[scale=0.5]{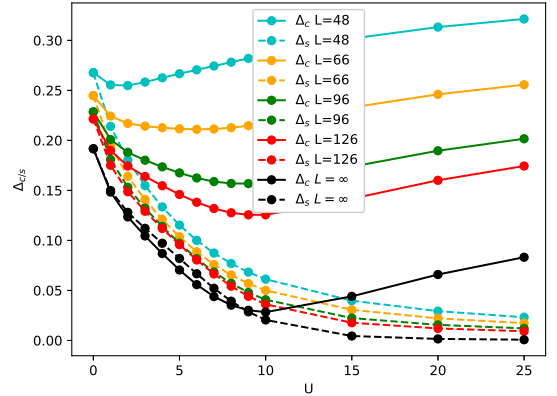} }
\par\end{centering}
\caption{Evolution of the charge and spin gaps with $U$, at filling $\nu=2/3$,
no magnetization and $Q=2\pi/3$. The spin gap decays smoothly towards
zero for large $U$, while the charge gap starts decaying up to $U\approx10\,t$
and then grows (below its band value) within the analyzed $U$ range.}
\label{fig: third filling m=00003D0 gaps}
\end{figure}

Ignoring electron interactions, the band structure in Fig. \ref{fig: third filling m=00003D0 spectrum}
shows equal charge and spin gaps. These gaps are modified by a repulsive
Hubbard interaction as shown in Fig. \ref{fig: third filling m=00003D0 gaps}.
According to the bosonized Hamiltonian in Eq. (\ref{eq:H-bos}) the
effect of $U$ is present because $k_{F,+}^{0}=k_{F,-}^{0}$ but there
is no Umklapp term as $2\left(k_{F,+}^{0}+k_{F,-}^{0}\right)=4\pi/3$
violates pseudo-momentum conservation; this provides a reason for
the similar behavior of the charge and spin gaps for moderate $U$,
up to $U\approx10\,t$. However, we find that the charge gap bounces
back for larger $U$\textcolor{blue}{{} }and then grows within the analyzed
$U$ range, remaining below its $U=0$ band value.

\begin{figure}
\begin{centering}
\includegraphics[scale=0.5]{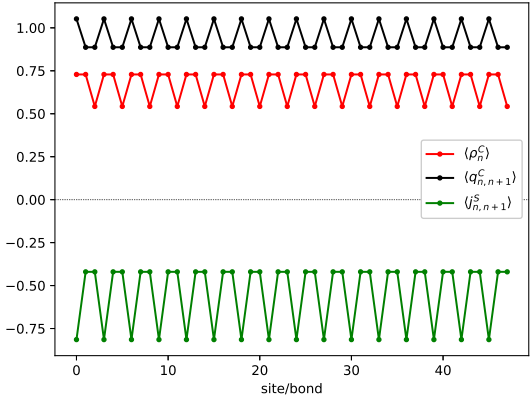} 
\par\end{centering}
\caption{Local oscillation of site and bond charge densities $\langle\rho_{n}^{C}\rangle$,
$\langle q_{n,n+1}^{C}\rangle$ and spin current density $\langle j_{n,n+1}^{S}\rangle$
in a wire of $L=48$ sites (PBC), with $Q=2\pi/3$, $\nu=2/3$ and
$m=0$. Local magnetization, bond spin density and charge current
density vanish (not shown).}
\label{fig: third filling m=00003D0 band order}
\end{figure}

\begin{figure}
\begin{centering}
\includegraphics[scale=0.5]{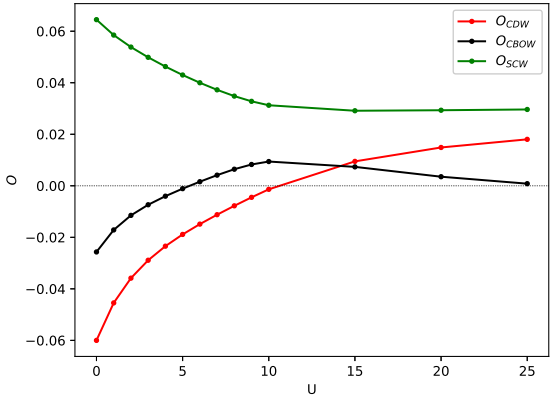} 
\par\end{centering}
\caption{Evolution of the non vanishing order parameters with $U$, for a wire
of $L=126$ sites at filling $\nu=2/3$ and no magnetization. }
\label{fig: third filling m=00003D0 orders}
\end{figure}

\textcolor{red}{}
Regarding order, this insulating phase shows period three oscillations
in the ground state expectation values of the site charge density
$\rho_{n}^{C}=\rho_{n}^{+}+\rho_{n}^{-}$, the bond charge density
$q_{n,n+1}^{C}$ and the spin current density $j_{n,n+1}^{S}$ as
shown in Fig. \ref{fig: third filling m=00003D0 band order}. The
evolution of the corresponding order parameters is shown in Fig. \ref{fig: third filling m=00003D0 orders}.
The charge bond order wave gets damped for large $U$ but the spin
current wave and the charge density wave stay present in this regime.
The latter gets out of phase with respect to RSOI modulations just
when the charge gap starts to increase. These findings might signal
a non-perturbative process that could be investigated in a future
work.

\section{Summary and conclusions}

In the present work we analyze the charge and spin order structure
in several gapped phases found in a quantum wire with spatially modulated
Rashba spin-orbit interaction, including electron-electron repulsive
interactions and a magnetic field along the spin-orbit axis. We classify
the conditions for the existence of a charge gap in absence of interactions
and provide numerical data for both the charge and spin gaps and a
set of proposed order parameters in the repulsive regime, obtained
by DMRG computations in finite length wires of up to $L=128$ electron
sites. The main features observed are supported by analytic arguments
within the bosonization framework. 

We first recall {[}\onlinecite{JJF_09}{]} that insulating phases,
namely ground states with a finite charge gap, can only be obtained
when the modulated Rashba coupling contains both uniform and oscillating
terms. In the present analysis we consider a single frequency modulation
$\gamma_{R}(n)=\gamma_{0}+\gamma_{1}\cos\left(Qn\right)$ with $\gamma_{0}\neq0$
and h $\gamma_{1}\neq0$ and provide commensurability conditions for
the charge gap opening, which relate the modulation wave number $Q$,
the electron filling and the magnetization {[}\onlinecite{Cabra_etal_17}{]}.

The emerging result in all of the possible insulator phases is the
presence of a long range order modulated spin current (spin current
wave), accompanied by a charge bond order wave. The charge gap and
the spin current wave are found to be robust under Hubbard electron-electron
interactions, proven up to $U=25\,t$. We relate this result to the
modified expression of particle current conservation under the evolution
dictated by the modulated Rashba Hamiltonian.

In the most prominent insulating phase of the system, that with particle
filling fraction one-half and no magnetic field, we show unexpected
symmetry properties such as particle-hole duality between the repulsive
and attractive regimes and the possibility of classifying quantum
states with charge $SU(2)$ quantum numbers, besides standard $SU(2)$
spin quantum numbers. These properties are explained by means of a
gauge mapping relating the modulated Rashba Hamiltonian with a modulated
hopping Hubbard Hamiltonian without spin orbit interaction {[}\onlinecite{Kaplan-1983}{]}. 

Regarding half-filled phases, the effective model for low carrier
density electron systems and large Coulomb repulsion {[}\onlinecite{Anderson-1959,Moriya-1960}{]}
relates the modulated Rashba interaction with modulation and anisotropy
in exchange couplings and modulated Dzyaloshinskii-Moriya couplings
in an effective $S=1/2$ Heisenberg spin chain. The large Hubbard
repulsion $U$ analyzed in the present problem might provide hints
to understand the behavior of those particular models of quantum spin
chains.
\begin{acknowledgments}
G.L.R. is grateful to M. Arlego, C. Lamas, A. Iucci and A. Lobos for
helpful discussions. G.I.J. acknowledges useful discussions with M.
Menteshashvili on early stages of the work. This work was partially
supported by CONICET (Grant No. PIP 2015-813), Argentina, and the
Shota Rustaveli Georgian National Science Foundation through the grant
N FR-19-11872.
\end{acknowledgments}

\bigskip{}

\section*{Appendix A: Some exact results for non interacting electrons with
modulated RSOI}

The Hamiltonians in Eq. (\ref{eq:H_tau}) are partially diagonalized
after a Fourier transformation 
\begin{equation}
d_{n,\tau}=\frac{1}{\sqrt{L}}\sum_{k=-\pi}^{\pi}e^{ikn}d_{k,\tau}\label{eq:Fourier}
\end{equation}
into momentum space. In order to decouple the excitation modes, in
the case of a periodic wire with length $L$ and RSOI modulations
with rational wave number $Q=2\pi r/p$ (compatible with $L$), it
is convenient to write the pseudo-momentum $k$ as $k=k_{0}+\nu K$
with $k_{0}$ in a reduced Brillouin zone $-\frac{\pi}{p}\leq k_{0}<\frac{\pi}{p}$,
an index $\nu=0,\cdots,p-1$ and $K=2\pi p/L$. The Hamiltonians then
take the form 
\begin{equation}
H_{\tau}=\sum_{k_{0}}\,\sum_{\nu,\nu'=0}^{p-1}M_{\nu\nu'}(k_{0,}\tau)d_{k_{0}+\nu K,\tau}^{\dagger}d_{k_{0}+\nu'K,\tau}^{\phantom{\dagger}}\,.\label{eq:H_pre-diagonal}
\end{equation}
which still couples sets of $p$ modes with pseudo-momenta differing
by $\nu K$. Diagonalization of the Hermitian $p\times p$ matrices
$M(k_{0},\tau)$ by means of unitary transformations $U(k_{0},\tau)$
unravels the elementary excitations 
\begin{equation}
H_{\tau}=\sum_{k_{0},\rho}\epsilon_{\rho}(k_{0},\tau)f_{k_{0},\rho,\tau}^{\dagger}f_{k_{0},\rho,\tau}^{\phantom{\dagger}}\label{eq:H_diagonal}
\end{equation}
where $\rho=0,\cdots,p-1$ is a band index, $\epsilon_{\rho}(k_{0},\tau)$
are the corresponding band dispersion relations and the fermionic
operators $f_{k_{0},\rho,\tau}^{\phantom{\dagger}}$ are related to
$d_{k,\tau}^{\phantom{\dagger}}$ in Eq. (\ref{eq:Fourier}) by 
\begin{equation}
d_{k_{0}+\nu K,\tau}^{\phantom{\dagger}}=\sum_{\rho}U_{\nu\rho}(k_{0},\tau)\,f_{k_{0},\rho,\tau}^{\phantom{}}.\label{eq:k-grouping}
\end{equation}

One can readily prove that the on-site order parameters in Eqs. (\ref{eq:O_CDW},
\ref{eq:O_SDW}) are expressed as \begin{widetext}
\begin{equation}
O_{CDW}=\frac{1}{L}\sum_{k_{0},\rho,\tau}\frac{1}{2}\sum_{j=\pm}\sum_{\nu}n_{\rho}(k_{0},\tau)U_{\nu+jr,\rho}^{*}(k_{0},\tau)U_{\nu\rho}(k_{0},\tau)\label{eq:O_CDW-ready}
\end{equation}
 and 
\begin{equation}
O_{SDW}=\frac{1}{L}\sum_{k_{0},\rho,\tau}\frac{\tau}{2}\sum_{j=\pm}\sum_{\nu}n_{\rho}(k_{0},\tau)U_{\nu+jr,\rho}^{*}(k_{0},\tau)U_{\nu\rho}(k_{0},\tau),\label{eq:O_SDW-ready}
\end{equation}
\end{widetext} where $n_{\rho}(k_{0},\tau)=\langle\tilde{f}_{k_{0},\rho,\tau}^{\dagger}\tilde{f}_{k_{0},\rho,\tau}^{\phantom{\dagger}}\rangle$
are the occupation numbers of states in the $\rho$-th band, with
momentum $k_{0}$ and spin $\tau$. For the zero temperature ground
state in a given filling and magnetization regime, these occupation
numbers are $0$ or $1$ according to the band structure and the corresponding
band filling fractions $\nu_{\tau}$. 

Bond-located order parameters in Eqs. (\ref{eq:O_CBOW}-\ref{eq:O_SCW})
can be recovered from the more general modulated average
\begin{equation}
O_{bond}^{\tau}=\frac{1}{L}\sum_{n}\cos(Qn)\langle d_{n,\tau}^{\dagger}d_{n+1,\tau}\rangle,\label{eq:O_bond}
\end{equation}
which after diagonalization reads \begin{widetext}
\begin{equation}
O_{bond}^{\tau}=\frac{1}{L}\sum_{k_{0},\rho}\frac{1}{2}\sum_{j=\pm}\sum_{\nu}n_{\rho}(k_{0},\tau)e^{i\left(k_{0}+\nu K\right)}U_{\nu+jr,\rho}^{*}(k_{0},\tau)U_{\nu\rho}(k_{0},\tau).\label{eq:O_bond-ready}
\end{equation}
\end{widetext}Then, being $2\,d_{n,\tau}^{\dagger}d_{n+1,\tau}=q_{n,n+1}^{\tau}-i\,j_{n,n+1}^{\tau}$,
one finds that
\begin{eqnarray}
O_{CBOW} & = & \sum_{\tau}2\,\mathbb{R}e(O_{bond}^{\tau}),\nonumber \\
O_{SBOW} & = & \sum_{\tau}2\tau\,\mathbb{R}e(O_{bond}^{\tau}),\label{eq:O_from_O_bond}\\
O_{CCW} & = & -\sum_{\tau}2\,\mathbb{I}m(O_{bond}^{\tau}),\nonumber \\
O_{SCW} & = & -\sum_{\tau}2\tau\,\mathbb{I}m(O_{bond}^{\tau}).\nonumber 
\end{eqnarray}

The $p\times p$ matrices $M_{\nu\nu'}(k_{0,}\tau)$ can be conveniently
diagonalized with the help of numerical routines, providing the energy
dispersion bands and the corresponding one-particle eigenstates. Then
the ground state for a given filling and magnetization, as well as
its order parameters, can be exactly computed. Still, we find it useful
to give analytical expressions for the simplest case $r=1$, $p=2$.
In this case $Q=\pi$ and one performs the diagonalization of 
\begin{equation}
M(k_{0},\tau)=\left(\begin{array}{cc}
-2\tilde{t}\cos(k_{0}-\tau q_{0}) & -i\tau2\gamma_{1}\cos(k_{0})\\
i\tau2\gamma_{1}\cos(k_{0}) & 2\tilde{t}\cos(k_{0}-\tau q_{0})
\end{array}\right)\label{eq:M_2x2}
\end{equation}
with $-\frac{\pi}{2}\leq k_{0}<\frac{\pi}{2}$, ignoring for the moment
diagonal terms proportional to the chemical potential and the magnetic
field. The energy bands, labeled by $\rho=0,\,1$, are found to be
$\epsilon_{\rho}(k_{0},\tau)=(-1)^{\rho+1}\epsilon_{\tau}(k_{0})$
with 
\begin{equation}
\epsilon_{\tau}(k_{0})=2\sqrt{\tilde{t}^{2}\cos^{2}(k_{0}-\tau q_{0})+\gamma_{1}^{2}\cos^{2}(k_{0})},\label{eq:dispersion_0}
\end{equation}
while the unitary matrices diagonalizing $M(k_{0},\tau)$ are given
by
\begin{equation}
U(k_{0},\tau)=\left(\begin{array}{cc}
\cos(\theta_{\tau}(k_{0})) & -i\tau\sin(\theta_{\tau}(k_{0}))\\
-i\tau\sin(\theta_{\tau}(k_{0})) & \cos(\theta_{\tau}(k_{0}))
\end{array}\right)\label{eq:U_2x2}
\end{equation}
with 
\begin{equation}
2\theta_{\tau}(k_{0})=\arctan\left(\frac{\gamma_{1}\cos(k_{0})}{\tilde{t}\cos(k_{0}-\tau q_{0})}\right)\,.\label{eq:theta}
\end{equation}

\section*{Appendix B. Conserved currents}

Consider a lattice Hamiltonian with time-independent coefficients
\begin{equation}
H=\frac{1}{2}\sum_{n}\left[\sum_{m\in N(n)}\cdots\right]\label{eq: Hamiltonian, general}
\end{equation}
for some local degrees of freedom (spins, bosons, fermions) where
$N(n)$ stands for the set of neighbor sites $m$ connected with $n$
by local interactions. Consider also a local density operator $\rho_{n}$
(for instance a local number operator). In the Heisenberg picture
we can write the time evolution of $\rho_{n}$ as 
\begin{equation}
\frac{d}{dt}\rho_{n}=i[H,\rho_{n}].\label{eq: time evolution}
\end{equation}
A continuity equation for $\rho_{n}$ in the lattice should relate
this time rate with the flow of local currents $J_{n\to m}$ transporting
density from the site $n$ into neighbors $m\in N(n)$. Thus we get
\begin{equation}
-\frac{d}{dt}\rho_{n}=\sum_{m\in N(n)}J_{n\to m}.\label{eq: current definition}
\end{equation}
From the actual form of the r.h.s. of Eq. (\ref{eq: current definition})
for a given model we can define current operators that describe the
flow of $\rho$ from site $n$ to site $m$. Notice that the expression
for current operators defined in this way depends not only on the
degrees of freedom involved, but also on the Hamiltonian structure
and coefficients. In a stationary state $|\psi\rangle$ one finds
that $\langle\psi|\rho_{n}|\psi\rangle$ does not evolve with time,
then the net current flow from each site vanishes
\begin{equation}
\sum_{m\in N(n)}\langle\psi|J_{n\to m}|\psi\rangle=0.\label{eq: generalized current conservation}
\end{equation}

Considering the Hamiltonian in Eq. (\ref{eq:H-bilinears}) and spin
polarized densities $\rho_{n,\tau}$ in Eq. (\ref{eq: density operator})
we find that
\begin{equation}
\frac{d}{dt}\rho_{n}^{\tau}=-i[H,\rho_{n}^{\tau}]=J_{n\to n+1}^{\tau}+J_{n\to n-1}^{\tau}\label{eq: rho_tau evolution}
\end{equation}
with 
\begin{equation}
J_{n\to n+1}^{\tau}=t\,j_{n,n+1}^{\tau}+\tau\gamma_{R}(n)\,q_{n,n+1}^{\tau}\label{eq: genearlizd current definition}
\end{equation}
mixing what we have called spin polarized bond density $q_{n,n+1}^{\tau}$
and spin polarized current $j_{n,n+1}^{\tau}$ in the main text (see
Eqs. (\ref{eq: q operator}, \ref{eq: j operator})). Eqs. (\ref{eq: generalized current conservation},
\ref{eq: rho_tau evolution}), together with inversion symmetry (w.r.t.
bond centered inversion points), show that at any bond 
\begin{equation}
\langle J_{n\to n+1}^{(\tau)}\rangle=0\,.\label{eq: current vanishes}
\end{equation}

\section*{Appendix C. Particle-hole duality and $SU(2)\times SU(2)$ symmetry}

In the main text, according to our interest, we have respected the
distinction between hopping terms and current terms. An alternative
strategy starts by grouping real and imaginary coefficients of $H$
in Eq. (\ref{eq:H-bilinears}) into complex coefficients as
\begin{eqnarray}
H & = & -\sum_{n,\tau}\left(\tilde{t}_{n}e^{-i\tau\phi_{n}}d_{n,\tau}^{\dagger}d_{n+1,\tau}^{\phantom{\dagger}}+H.c\right)\nonumber \\
 & - & \sum_{n,\tau}\left(\mu+\tau\frac{h_{y}}{2}\right)\rho_{n,\tau}\label{eq:H_tau-1}\\
 & + & U\sum_{n}\rho_{n,+}\rho_{n,-}\,\nonumber 
\end{eqnarray}
where $\tilde{t}_{n}=\sqrt{t^{2}+\gamma_{R}^{2}(n)}$ and $\tan\phi_{n}=\gamma_{R}(n)/t$.
One can then perform {[}\onlinecite{Kaplan-1983}{]} the following
gauge transformation 
\begin{align}
d_{n,\tau}^{\phantom{\dagger}} & \to e^{i\tau\theta_{n}}d_{n,\tau}^{\phantom{\dagger}}\label{eq: gauge trafo}
\end{align}
with 
\begin{equation}
\theta_{n}=\sum_{m<n}\phi_{m}\label{eq: accumulated phas}
\end{equation}
so that 
\begin{equation}
d_{n,\tau}^{\dagger}d_{n+1,\tau}^{\phantom{\dagger}}\to e^{i\tau\phi_{n}}d_{n,\tau}^{\dagger}d_{n+1,\tau}^{\phantom{\dagger}}\label{eq: bilinear trafo}
\end{equation}
while densities $\rho_{n,\tau}=d_{n,\tau}^{\dagger}d_{n,\tau}^{\phantom{\dagger}}$
remain invariant. In this way $H$ is mapped onto a Hubbard model
with (real) modulated hopping coefficients and no RSOI interactions.
It appears appropriate to say that the RSOI is gauged away by the
above procedure. Notice that the transformation in Eq. (\ref{eq: gauge trafo})
can be also depicted as local spinor rotations around the $\hat{y}$
axis, 
\begin{equation}
d_{n}\to e^{i\sigma_{y}\theta_{n}}d_{n}\label{eq: local rotation}
\end{equation}
where $d_{n}^{\dagger}=\left(d_{n,+}^{\dagger},d_{n,-}^{\dagger}\right)$
stands for the fermionic operators in spinor form and $\vec{\sigma}$
are the Pauli matrices (rotated in order to make $\sigma_{y}=diag(1,-1)$,
see Eq. (\ref{spinor})); this immediately allows for writing the
mapping as a unitary transformation ${\cal U}$ in the Hilbert space,
$H\to H_{\text{Hubbard}}={\cal U}^{-1}H\,{\cal U}$. A related transformation
has been presented in earlier works {[}\onlinecite{Perk-1976,Kaplan-1983}{]},
and recently used in {[}\onlinecite{AJR_19}{]} to gauge away spin-orbit
interactions from spin chain models. 

The existence of the mapping between $H$ in Eq. (\ref{eq:H-bilinears})
and a modulated hopping Hubbard model reveals a hidden $SU(2)$ spin
symmetry. To be explicit, as the Hubbard Hamiltonian has the usual
global $SU(2)$ symmetry generated by the spin operators
\begin{equation}
\vec{J}=\sum_{n}\frac{1}{2}d_{n}^{\dagger}\vec{\sigma}d_{n}^{\phantom{\dagger}},\label{eq: SU(2) spin}
\end{equation}
then the modulated RSOI Hamiltonian has a symmetry $[{\cal U}\,\vec{J}\,{\cal U}^{-1},H]=0$
with \char`\"{}twisted\char`\"{} spin generators ${\cal U}\,\vec{J}\,{\cal U}^{-1}$
satisfying an $SU(2)$ algebra. These \char`\"{}twisted\char`\"{}
spin generators are a generalization of those discussed in {[}\onlinecite{Bernevig-2006}{]}.

Moreover, in the half-filled non magnetized phase discussed in Section
\ref{sec:Non-magnetized-insulator-half-filling} the target Hamiltonian
$H_{\text{Hubbard}}$ possesses enhanced symmetries: it is particle-hole
dual and invariant under $SU(2)\times SU(2)$ spin and charge transformations
{[}\onlinecite{Yang-1990-1991}{]}. Indeed, in that Hubbard case the
particle-hole transformation {[}\onlinecite{Shiba-1972}{]} is given
by
\begin{equation}
\begin{cases}
d_{n,+}^{\phantom{\dagger}}\to & d_{n,+}^{\phantom{\dagger}}\\
d_{n,-}^{\phantom{\dagger}}\to & (-1)^{n}d_{n,-}^{\dagger}
\end{cases}\label{eq: particle-hole trafo}
\end{equation}
and can be implemented as a unitary transformation ${\cal X}$ with
the duality property 
\begin{equation}
{\cal X}\,H_{\text{Hubbard}}(U)\,{\cal X}^{-1}=H_{\text{Hubbard}}(-U).\label{eq: Hubbard duality}
\end{equation}
The charge $SU(2)$ generators for the Hubbard model are obtained
as a particle-hole transformation of the spin generators, $\vec{J}_{\text{charge}}={\cal X}\,\vec{J}\,{\cal X}^{-1}$.
They are proven to satisfy the $SU(2)$ algebra, to commute with $H_{\text{Hubbard}}$
and also to commute with $\vec{J}$, so that $(\vec{J},\vec{J}_{\text{charge}})$
generate an extended $SU(2)\times SU(2)$ symmetry. Mapping back these
generators onto the problem in Section \ref{sec:Non-magnetized-insulator-half-filling}
one learns that ${\cal U}\,{\cal X}\,{\cal U}^{-1}$ is a particle-hole
duality transformation and that $\left({\cal U}\,\vec{J}\,{\cal U}^{-1},\,{\cal U}\,\vec{J}_{\text{charge}}\,{\cal U}^{-1}\right)$
generate a $SU(2)\times SU(2)$ symmetry on the RSOI Hamiltonian $H$.
As the mapping preserves total charge and total magnetization along
the $\hat{y}$ axis, one can identify charge and spin sectors of both
models.

As we have seen, gauging away the RSOI brings theoretical insight
into the problem of interest in the present work. However, it has
the cost of introducing a twist, and the associated numerical difficulties,
in the boundary conditions for finite size chains {[}\onlinecite{Goth_2014}{]}.

\end{document}